\documentclass[11pt]{amsart}

\usepackage[margin=1.15in]{geometry}
\usepackage{amsmath,amssymb,amsthm,mathtools}
\usepackage{enumitem}
\usepackage{xcolor}
\usepackage{microtype}
\usepackage[colorlinks=true,linkcolor=blue!50!black,citecolor=blue!50!black,urlcolor=blue!50!black]{hyperref}
\usepackage{graphicx}
\usepackage{float}

\newtheorem{theorem}{Theorem}[section]
\newtheorem{proposition}[theorem]{Proposition}
\newtheorem{lemma}[theorem]{Lemma}
\newtheorem{conjecture}[theorem]{Conjecture}
\theoremstyle{definition}
\newtheorem{definition}[theorem]{Definition}
\theoremstyle{remark}
\newtheorem{remark}[theorem]{Remark}

\numberwithin{equation}{section}

\newcommand{\R}{\mathbb{R}}
\newcommand{\Z}{\mathbb{Z}}
\newcommand{\E}{\mathbb{E}}
\renewcommand{\Pr}{\mathbb{P}}
\newcommand{\cx}{\preceq_{\mathrm{cx}}}
\newcommand{\one}{\mathbf{1}}

\newcommand{\Wit}{\mathcal{W}_D}

\newcommand{\slope}{\operatorname{sl}}
\DeclareMathOperator{\supp}{supp}

\begin{document}

\title{Online carpooling against an oblivious adversary}

\author{Yuhao Lei}
\address{Department of Mathematical Sciences, Carnegie Mellon University, Pittsburgh, PA 15213, USA}
\email{yuhaolei@andrew.cmu.edu}

\begin{abstract}
In the online carpooling problem, edges on $n$ vertices arrive one at a time and must be oriented immediately, so as to keep the difference between the indegree and the outdegree of every vertex small.
We give a randomized online algorithm which bounds the largest absolute difference over the first $T$ edges by $O((\log T)^{1/3})$ with high probability on every input sequence fixed in advance. This proves a conjecture of Kulkarni, Reis and Rothvoss and matches the known worst case lower bound. The algorithm keeps the law of the discrepancy vector dominated in the convex order by a Gaussian measure, and forces the endpoint with the smaller discrepancy to receive the edge whenever the two discrepancies are far apart. In the analysis we show that a vertex of discrepancy $D$ is preceded by a set of $O(D)$ vertices with total discrepancy $\Omega(D^2)$. 
\end{abstract}
\maketitle

\section{Introduction}

\subsection{Online carpooling and edge orientation}

Let $n\ge 2$ and $[n]=\{1,\dots,n\}$. In the online carpooling problem, edges arrive one at a time, and the $t$-th edge joins two distinct vertices $u_t$ and $v_t$ of $[n]$. When an edge arrives, an online algorithm must orient it, either from $u_t$ to $v_t$ or from $v_t$ to $u_t$, without seeing the later edges and without revising earlier decisions. We call the endpoint whose discrepancy increases the receiver and the other endpoint the sender. The aim is to keep, at all times, the difference between the indegree and the outdegree of every vertex small. In the pairwise carpooling model inspired by Fagin and Williams~\cite{FW83}, vertices represent people, each edge represents a shared ride, and its orientation determines the driver. The discrepancy of each vertex measures the deviation of that person's driving count from its fair share.

It is convenient to state the problem as an online vector balancing problem. For $x\in\R^n$ we write $x(i)$ for its $i$-th coordinate, $\|x\|$ for its Euclidean norm and $\|x\|_\infty=\max_i|x(i)|$, and $\langle\cdot,\cdot\rangle$ denotes the standard inner product on $\R^n$. Let $\one_i$ be the $i$-th standard basis vector of $\R^n$. We represent the $t$-th edge by $e_t=\one_{u_t}-\one_{v_t}$ and also call this vector an edge. The algorithm chooses a sign $\varepsilon_t\in\{-1,+1\}$; the sign $+1$ means that the edge is oriented from $v_t$ to $u_t$. The discrepancy vector evolves by
\begin{equation}\label{eq:walk}
d_0=0,\qquad d_t=d_{t-1}+\varepsilon_te_t\quad(t\ge 1).
\end{equation}
Thus $d_t(i)$ is the indegree minus the outdegree of the vertex $i$ in the graph formed by the first $t$ oriented edges. The quantity of interest is
\begin{equation}\label{eq:MT}
M_T=\max_{0\le t\le T}\|d_t\|_\infty .
\end{equation}

An \emph{adaptive} adversary chooses $e_t$ after seeing $\varepsilon_1,\dots,\varepsilon_{t-1}$. An \emph{oblivious} adversary fixes the whole sequence before the algorithm starts, so that the edges do not depend on the random choices of the algorithm.

This paper is about the oblivious setting. All probabilities and expectations are over the internal randomness of the algorithm. A \emph{run} is a realization of this randomness, and its \emph{trajectory} is the resulting discrepancy sequence $(d_t)_{0\le t\le T}$.

\subsection{Background and main result}
Online vector balancing goes back to the balancing games of Spencer~\cite{Spe77}. The subject has developed along several directions: stochastic arrivals where the input vectors are sampled from a distribution, adaptive adversary, and worst case arrivals fixed by an oblivious adversary. For independent arrivals from an arbitrary distribution on $[-1,1]^n$, Bansal, Jiang, Singla and Sinha~\cite{BJSS20} obtained prefix discrepancy bounds polynomial in $n$ and $\log T$. Their framework handles dependence among the coordinates of an arriving vector and also yields bounds for geometric discrepancy problems. For independent uniformly random vectors in $\{-1,+1\}^n$, Bansal and Spencer~\cite{BS20} achieved the optimal $O(\sqrt n)$ terminal discrepancy with high probability. The subsequent development for stochastic arrivals by Bansal, Jiang, Meka, Singla and Sinha~\cite{BJMSS21} developed a potential based approach that improved these bounds and extended the results to more general discrepancy settings.

Alweiss, Liu and Sawhney~\cite{ALS21} constructed a randomized online algorithm that, against an oblivious adversary, gave discrepancy bounds optimal up to polylog. Their self-balancing walk chooses each sign at random, biased bias toward the update
that decreases the Euclidean norm of the discrepancy vector. Liu, Sah and Sawhney~\cite{LSS22} developed a closely related approach through random walks with a Gaussian stationary distribution. A significant advance of Kulkarni, Reis and Rothvoss~\cite{KRR24} established the optimal general bound $O(\sqrt{\log T})$ for vectors of Euclidean norm at most one, the online Koml\'os setting. They constructed an online algorithm for which every fixed prefix sum is subgaussian with an absolute parameter, and proved a matching lower bound against an oblivious adversary. Aden-Ali~\cite{AA26} subsequently provided an efficient algorithm with the same discrepancy order in $O(nT)$ time by coupling three Gaussian fixed point walks so that every update uses a sign in $\{-1,+1\}$. Recent work also identifies how sparsity changes the problem. For vectors in $[-1,1]^n$ with at most $d$ nonzero coordinates, Altschuler and Tikhomirov~\cite{AT26a} obtained an efficient online algorithm with $O(\sqrt d)$ prefix discrepancy down to nearly logarithmic sparsity: for any fixed $\eta>0$, their bound applies when $d\ge C_\eta(\log T)(\log\log T)^{2+\eta}$. For $T=\Theta(n)$ independent uniformly random binary vectors with exactly $d$ nonzero coordinates, they determined the optimal prefix discrepancy order $\Theta\bigl(\max\{\sqrt d,\log\log n\}\bigr)$ when \(2\le d\le n/2\)
in~\cite{AT26b}. Carpooling has fixed sparsity two and the additional structure that its two nonzero coordinates have opposite signs. For broader background on the algorithmic and geometric aspects of discrepancy, we refer to the survey of Bansal~\cite{Ban23} and the monograph of Bansal and Nikolov~\cite{BN26}.

For carpooling, against adaptive adversaries the optimal order is known. Ajtai, Aspnes, Naor, Rabani, Schulman and Waarts~\cite{AAN98} showed that an adaptive adversary can force discrepancy $\Omega(\min\{T^{1/3},n\})$, and Bansal, Prabhu, Singla and Sundaram~\cite[Theorem~1.1]{BPSS26} proved the matching upper bound for the greedy algorithm, which orients each edge
towards the endpoint of smaller discrepancy. They proved that for $T=O(n)$ edges sampled independently and uniformly from a $\Delta$-regular graph, discrepancy $O((\log n)^{1/3})$ is achievable with high probability, and asked whether $O((\log T)^{1/3})$ holds for edges sampled from an arbitrary graph~\cite[Section~1.1]{BPSS26}. For related work on Greedy and on random inputs see~\cite{CNP11,FKK16,GKKS20}.

Randomization helps against oblivious adversaries. A randomized
algorithm of Kalai, reported by Naor~\cite{Nao05} gives $M_T=O(\log T)$ with high probability. Applying the general algorithms of~\cite{KRR24,AA26} to the normalized edge vectors $e_t/\sqrt2$ improves this to $O(\sqrt{\log T})$: each fixed time coordinate is subgaussian. A lower bound of order $(\log T)^{1/3}$ is implicit in~\cite[Section~3.2]{AAN98}; see also~\cite[Section~1]{BPSS26} and~\cite{FKK16}. This is a worst case lower bound in which the number of vertices may grow with steps. Kulkarni, Reis and Rothvoss~\cite[Conjecture~2]{KRR24} conjectured that a randomized online algorithm achieves the matching upper bound on every fixed edge sequence.

\begin{conjecture}[{\cite[Conjecture~2]{KRR24}}]\label{conj:KRR}
There is a randomized online algorithm for carpooling which, for every $T\ge 1$ and every sequence of $T$ edges fixed in advance, satisfies $M_T=O((\log T)^{1/3})$ with high probability.
\end{conjecture}

The goal of this paper is to prove this conjecture.

\begin{theorem}\label{thm:main}
For every $n\ge2$ there is a randomized online algorithm, independent of $T$, such that for every sequence of $T\ge1$ edges on $[n]$ fixed in advance and every integer $D\ge300$, the maximum $M_T$ defined in~\eqref{eq:MT} satisfies
\begin{equation}\label{eq:main-tail}
 \Pr(M_T\ge D)\le12T\exp\left[-\left(\frac D{40}\right)^3\right].
\end{equation}
Consequently, for every $c\ge0$, with probability at least $1-e^{-c}$,
\begin{equation}\label{eq:main-whp}
 M_T<\max\left\{300,\;40\bigl(c+\log(12T)\bigr)^{1/3}\right\}.
\end{equation}
\end{theorem}

In particular, this answers the question in~\cite[Section~1.1]{BPSS26} for arbitrary graphs, with no regularity assumption or restriction on the time horizon.

\subsection{Ideas of the proof}\label{sec:ideas}
The geometric starting point is Banaszczyk's vector balancing theorem~\cite{Ban98} and the convex set transform in its proof. For a sufficiently short balancing vector, this transform does not decrease the Gaussian measure of a convex set whose measure is at least one half. This (sub)gaussian point of view also underlies the advances in~\cite{KRR24,LSS22,AA26} where the algorithms of~\cite{KRR24,AA26} provide the $O(\sqrt{\log T})$ discrepancy bound. To improve on $O(\sqrt{\log T})$ we consider backward witness, a variant of the spreadness argument of the recent advance ~\cite[Lemma~3.3]{BPSS26} whose witnesses are determined by the input and $O(D^2)$ ternary choices. A vertex of discrepancy $D$ leads backwards to a set $S$ of $O(D)$ vertices carrying total discrepancy $\Omega(D^2)$. Subgaussianity bounds the probability of this by $\exp(-\Omega(D^3))$ so that a union bound over them gives the order $(\log T)^{1/3}$ against an oblivious adversary. 

We construct an algorithm in Section~\ref{sec:algorithm} with the following two properties:
\begin{enumerate}[label=(P\arabic*),leftmargin=3.2em]
\item after each edge is oriented, the receiver is ahead of the sender by at most a constant;
\item at every fixed time, $d_t$ is subgaussian with a constant parameter.
\end{enumerate}
Greedy has (P1), with the constant $2$, but not (P2); the algorithms of~\cite{KRR24,AA26} have (P2) but are not known to have (P1). Without (P1), a single vertex can grow by receiving from vertices far below it: if $n\ge T+1$, the edges $\one_1-\one_{t+1}$, $1\le t\le T$, arrive and the vertex $1$ always receives, then its discrepancy reaches $T$, while no other vertex is ever above $0$.

Section~\ref{sec:witness} derives $\Pr(M_T\ge D)\le O(T)\exp(-\Omega(D^3))$ from (P1) and (P2). Starting from a vertex $i^*$ that reaches discrepancy $D$ at the first such time $t^*$, we walk backwards with a set $S$, initially $\{i^*\}$, and compare $d_t(S)=\sum_{i\in S}d_t(i)$ with a threshold $A(|S|)$. At an edge with exactly one endpoint in $S$, we keep $S$, add the outside endpoint or remove the inside one. These choices form a ternary tree determined by the input and the root $(t^*,i^*)$. Property (P1) ensures that one of the choices raises $d_t(S)-A(|S|)$ by at least one. Following such choices leads to a set with total discrepancy $\Omega(D^2)$ once $|S|$ is a fixed fraction of $D$ or after $O(D^2)$ crossings. We stop every path after at most $O(D^2)$ branching steps, giving $\exp(O(D^2))$ leaves per root. Since $i^*$ is an endpoint of $e_{t^*}$, there are at most $2T$ roots. The resulting family of leaves depends only on the input, and the union bound gives
\[
\Pr(M_T\ge D)\le\underbrace{T\exp(O(D^2))}_{\text{number of candidates}}\cdot\underbrace{\exp(-\Omega(D^3))}_{\text{bound from (P2)}} .
\]

The algorithm in Section~\ref{sec:algorithm} keeps the law $\mu_t$ of $d_t$ dominated in convex order by the centered Gaussian measure $\gamma$ on $\R^n$ with covariance $50$ times the identity; exponential test functions then give (P2). The measure $\mu_{t-1}$ is determined recursively by $e_1,\dots,e_{t-1}$, and it is the law of $d_{t-1}$. When $e_t$ arrives, the algorithm chooses the probability of moving from $x$ to $x+e_t$ rather than to $x-e_t$ for each point $x$ of the support of $\mu_{t-1}$ such that the new law is again dominated by $\gamma$ and the endpoint that is behind receives whenever $|x(u_t)-x(v_t)|>27$; this gives (P1) with the constant $29$. Such a choice exists by Theorem~\ref{thm:selection} which is proved in two steps in Sections~\ref{sec:onedim} and~\ref{sec:selection}.

\section{Preliminaries}\label{sec:prelim}

For $S\subseteq[n]$ write $\one_S=\sum_{i\in S}\one_i$ and $x(S)=\sum_{i\in S}x(i)=\langle x,\one_S\rangle$; thus $\|\one_S\|^2=|S|$. 

We write $\delta_x$ for the unit mass at $x$ and $\supp\mu$ for the support of a measure $\mu$. 

An edge $e=\one_u-\one_v$ has $\|e\|^2=2$ and $\langle x,e\rangle=x(u)-x(v)$. It \emph{crosses} $S$ if exactly one of its endpoints belongs to $S$.

Time $t$ refers to the state after the $t$-th edge has been oriented. By~\eqref{eq:walk}, $d_t\in\Z^n$, and at step $t$ the two endpoint discrepancies change by one in opposite directions. 

Define the receiver $a_t$ and sender $b_t$ by $\varepsilon_te_t=\one_{a_t}-\one_{b_t}$, and put $L_t=\langle d_{t-1},e_t\rangle$. Then
\begin{equation}\label{eq:lead}
d_t(a_t)-d_t(b_t)
=\langle d_t,\varepsilon_te_t\rangle
=2+\varepsilon_tL_t.
\end{equation}
The \emph{receiving gap} of a trajectory is
\[
\max_{1\le t\le T}\bigl(d_t(a_t)-d_t(b_t)\bigr).
\]

Let $\gamma=N(0,50I_n)$ and let $G$ have law $\gamma$. For every $\theta\in\R^n$,
\[
\E e^{\langle\theta,G\rangle}=e^{25\|\theta\|^2};
\]
in particular, $\langle G,e\rangle\sim N(0,100)$ for every edge $e$.

\section{The algorithm}\label{sec:algorithm}

\subsection{Convex order and admissible selections}\label{sec:admissible}

Let $\mathcal C_+$ be the finite nonnegative convex functions on $\R^n$ that are integrable under $\gamma$.

\begin{definition}\label{def:cx}
For a finitely supported probability measure $\mu$, write $\mu\cx\gamma$ if
$\int f\,d\mu\le\int f\,d\gamma$ for every $f\in\mathcal C_+$.
\end{definition}

This is the usual convex order. For a finite convex function $f$ integrable under $\gamma$, apply the inequality to $(f+m)_+$ and subtract $m$. Since $(f+m)_+-m=\max\{f,-m\}$ and $|\max\{f,-m\}|\le|f|$, dominated convergence gives the inequality for $f$ as $m\to\infty$; integrability under $\mu$ follows from its finite support.

For a fixed edge $e$, a \emph{selection} for $\mu$ is a function $p\colon\supp\mu\to[0,1]$, identified with an element of $[0,1]^{\supp\mu}$. Its value $p(x)$ is the probability of moving from $x$ to $x+e$ rather than to $x-e$. Define
\begin{equation}\label{eq:mu-p}
\mu^{p,e}=\sum_{x\in\supp\mu}\mu(\{x\})\Big(p(x)\,\delta_{x+e}+\big(1-p(x)\big)\,\delta_{x-e}\Big).
\end{equation}

For $x\in\R^n$ let
\begin{equation}\label{eq:Bx}
B_x=\begin{cases}\{0\},&\langle x,e\rangle>27,\\ \{1\},&\langle x,e\rangle<-27,\\ [0,1],&|\langle x,e\rangle|\le27,\end{cases}
\end{equation}
and let $\mathcal B=\prod_{x\in\supp\mu}B_x$. These \emph{forcing conditions} make the endpoint with smaller discrepancy receive whenever the two discrepancies differ by more than $27$. Each $B_x$ is a nonempty compact interval, so $\mathcal B$ is nonempty, compact and convex.

\begin{definition}\label{def:admissible}
A selection $p\in\mathcal B$ is \emph{admissible} for $(\mu,e)$ if $\mu^{p,e}\cx\gamma$.
\end{definition}

\begin{theorem}\label{thm:selection}
Let $u\ne v$ be elements of $[n]$, put $e=\one_u-\one_v$, and let $\mu$ be a finitely supported probability measure on $\R^n$ with $\mu\cx\gamma$. Then the set of admissible selections for $(\mu,e)$ is a nonempty, compact and convex subset of $\mathcal B$.
\end{theorem}

Theorem~\ref{thm:selection} is proved in Section~\ref{sec:selection}.
As in~\cite{KRR24} where the algorithm is derived from Banaszczyk's transform of convex sets by an existence argument, the selection here is obtained nonconstructively, and we are interested to learn whether it admits a polynomial-time implementation.
The Gaussian fixed point walk of Liu, Sah and Sawhney~\cite{LSS22} is a walk on the line with steps $0$ and $\pm1$ whose stationary law is exactly Gaussian.
Here the law of $d_t$ is only dominated in convex order by $\gamma$, and this inequality leaves room for the forced moves.

\subsection{Definition of the algorithm}

Fix $n\ge2$. For every finite edge prefix we define probability measures $\mu_t$ and admissible selections $p_t$ recursively. Put $\mu_0=\delta_0$, which satisfies $\mu_0\cx\gamma$ by Jensen's inequality. Given $\mu_{t-1}$ and the next edge $e_t$, let $p_t$ minimize
\[
\sum_{x\in\supp\mu_{t-1}}\big(p(x)-\tfrac12\big)^2
\]
over the admissible selections for $(\mu_{t-1},e_t)$, and put $\mu_t=\mu_{t-1}^{p_t,e_t}$. Theorem~\ref{thm:selection} and strict convexity give a unique minimizer. Each $\mu_t$ is finitely supported and satisfies $\mu_t\cx\gamma$, and the recursion determines $\mu_t$ and $p_t$ from $e_1,\dots,e_t$.

Let $U_1,U_2,\ldots$ be independent uniform variables on $[0,1)$. At step $t$, set
\begin{equation}\label{eq:rule}
\varepsilon_t=\begin{cases}+1,& U_t<p_t(d_{t-1}),\\-1,&\text{otherwise,}\end{cases}
\end{equation}
and update $d_t$ by~\eqref{eq:walk}. For every realization of $(U_t)_{t\ge1}$, $d_t\in\supp\mu_t$: inductively, a chosen $+$ move has $p_t(d_{t-1})>0$, and a chosen $-$ move has $p_t(d_{t-1})<1$, so~\eqref{eq:mu-p} assigns positive mass to the next state. 

\begin{proposition}\label{prop:alg}
Let the input $e_1,e_2,\dots$ be fixed in advance.
Each trajectory has receiving gap at most $29$ and $d_t$ has law $\mu_t$. In particular,
$\E e^{\langle\theta,d_t\rangle}\le e^{25\|\theta\|^2}$ for all $t\ge0$ and $\theta\in\R^n$.
\end{proposition}
\begin{proof}
If $|L_t|>27$, the forcing conditions give $\varepsilon_tL_t=-|L_t|$; otherwise $\varepsilon_tL_t\le27$. Equation~\eqref{eq:lead} bounds the receiving gap.

The law identity holds at $t=0$. Suppose it holds at $t-1$. Since the input is fixed, $d_{t-1}$ depends only on $U_1,\dots,U_{t-1}$, so $U_t$ is independent of $d_{t-1}$. Conditional on $d_{t-1}=x$, the two moves therefore have probabilities $p_t(x)$ and $1-p_t(x)$. Formula~\eqref{eq:mu-p} gives $\mathcal L(d_t)=\mu_{t-1}^{p_t,e_t}=\mu_t$. Finally, $x\mapsto\exp(\langle\theta,x\rangle)$ belongs to $\mathcal C_+$, so convex order gives
\[
\E\exp\langle\theta,d_t\rangle
=\int\exp(\langle\theta,x\rangle)\,d\mu_t(x)
\le\int\exp(\langle\theta,x\rangle)\,d\gamma(x)
=\exp(25\|\theta\|^2).\qedhere
\]
\end{proof}

\section{Backward witnesses}\label{sec:witness}
The idea is influenced by~\cite{BPSS26}. 
We first construct a family of small sets at fixed times from the edges alone. Every trajectory with receiving gap at most $29$ and $M_T\ge D$ has total discrepancy of absolute value at least $D^2/80$ on one of these sets. We then apply Proposition~\ref{prop:alg} to each fixed candidate and sum over the family. In Sections~\ref{sec:staircase}--\ref{sec:backward}, fix the edges $e_1,\dots,e_T$ and an integer $D\ge300$.

\subsection{The threshold and the backward step}\label{sec:staircase}
Put
\begin{equation}\label{eq:params}
K=\Big\lfloor\frac{D}{40}\Big\rfloor,\qquad \ell=\Big\lceil\frac{D^2}{80}\Big\rceil,
\end{equation}
and, for integers $k\ge0$,
\begin{equation}\label{eq:staircase}
A(k)=\sum_{r=0}^{k-1}(D-30r)=kD-15k(k-1).
\end{equation}
The threshold $A$ has $A(0)=0$, $A(1)=D$, and increments $A(k+1)-A(k)=D-30k$. Bansal, Prabhu, Singla and Sundaram~\cite[Lemma~3.3]{BPSS26} compare the sum of the $k$ largest coordinates with $k(D-k+1)$, whose increments decrease by $2$, the receiving gap of Greedy. Here the increments decrease by $30$, one more than the upper bound $29$ on the receiving gap.

For $1\le k\le K$, since $30k \le 3D/4$  and $K> D/40-1\ge D/50$ we have
\begin{equation}\label{eq:staircase-lower}
A(k)\ge \tfrac58 kD>0,
\qquad
A(K)\ge \tfrac58 KD\ge \frac{D^2}{80}.
\end{equation}

For a trajectory $(d_t)_{0\le t\le T}$ and a set $S\subseteq[n]$, define the \emph{slack}
\[
\Lambda(t,S)=d_t(S)-A(|S|).
\]
When $e_t$ crosses $S$, write
\[
S^0=S,\qquad S^+=S\cup\{u_t,v_t\},\qquad S^-=S\setminus\{u_t,v_t\}.
\]
Thus $S^+$ adds the outside endpoint and $S^-$ removes the inside endpoint.

\begin{figure}[H]
\centering
\includegraphics[width=0.96\textwidth]{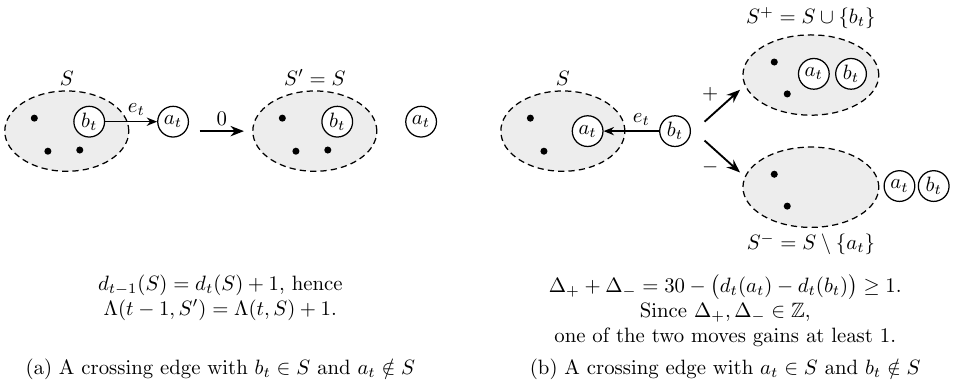}
\caption{The backward step. In (a) the slack increases by $1$; in (b) the two integer gains sum to at least $1$. The shaded regions indicate set membership.}
\label{fig:backward-step}
\end{figure}
If the sender lies in $S$, the slack rises in backtracking. Otherwise, adding the sender gains if its discrepancy exceeds $D-30|S|$, and removing the receiver gains if its discrepancy is below $D-30(|S|-1)$; these thresholds differ by $30$, more than the upper bound $29$ on the receiving gap, so one of them gains. The slack increases at every crossing during backtracking.
\begin{lemma}\label{lem:step}
Let $(d_t)_{0\le t\le T}$ have receiving gap at most $29$. If $1\le t\le T$, $S\subseteq[n]$ is nonempty, and $e_t$ crosses $S$, then one of the sets $S'\in\{S^0,S^+,S^-\}$ satisfies
\[
\Lambda(t-1,S')\ge\Lambda(t,S)+1.
\]
If moreover $\Lambda(t,S)\ge0$, then this choice has $S'\ne\emptyset$.
\end{lemma}

\begin{proof}
Since $e_t$ crosses $S$, exactly one of the receiver $a_t$ and the sender $b_t$ lies in $S$.

\emph{Case (i): $b_t\in S$ and $a_t\notin S$.} Take $S'=S^0=S$. Of the coordinates in $S$, only $b_t$ changes at step $t$, and $d_{t-1}(b_t)=d_t(b_t)+1$. Hence
\[
\Lambda(t-1,S)-\Lambda(t,S)=d_{t-1}(S)-d_t(S)=1.
\]

\emph{Case (ii): $a_t\in S$ and $b_t\notin S$.} In this case $S^+=S\cup\{b_t\}$ and $S^-=S\setminus\{a_t\}$, so $|S^+|=|S|+1$ and $|S^-|=|S|-1$. Both endpoints of $e_t$ lie in $S^+$ and neither lies in $S^-$, giving $d_{t-1}(S^\pm)=d_t(S^\pm)$. By~\eqref{eq:staircase},
\begin{align*}
\Delta_+&\coloneqq\Lambda(t-1,S^+)-\Lambda(t,S)=d_t(b_t)-\big(A(|S|+1)-A(|S|)\big)=d_t(b_t)-D+30|S|,\\
\Delta_-&\coloneqq\Lambda(t-1,S^-)-\Lambda(t,S)=-d_t(a_t)+\big(A(|S|)-A(|S|-1)\big)=-d_t(a_t)+D-30(|S|-1).
\end{align*}
Since the receiving gap is at most $29$,
\begin{equation}\label{eq:sum-of-gains}
\Delta_++\Delta_-=30-\big(d_t(a_t)-d_t(b_t)\big)\ge30-29=1.
\end{equation}
Both gains are integers. If $\Delta_+\ge1$, take $S'=S^+$. Otherwise $\Delta_+\le0$, so $\Delta_-\ge1-\Delta_+\ge1$, and we take $S'=S^-$.

In both cases $\Lambda(t-1,S')\ge\Lambda(t,S)+1$. If $S'=\emptyset$, then $\Lambda(t-1,S')=-A(0)=0$, so $\Lambda(t,S)\le-1$. Hence $S'\ne\emptyset$ whenever $\Lambda(t,S)\ge0$.
\end{proof}

\subsection{Witness tree}\label{sec:backward}
Since Lemma~\ref{lem:step} guarantees an improving backward step, but
the choice may depend on the discrepancy trajectory. We therefore retain all three possibilities
at each crossing edge; the resulting tree depends only on the edge sequence and $D$.

The reader may compare Lemma~\ref{lem:witness} with the spreadness lemma of~\cite[Lemma~3.3]{BPSS26}. Related witness tree arguments also
appear in graphical load balancing~\cite{KP06}.
\begin{lemma}\label{lem:witness}
There is a family $\Wit$ of pairs $(t,S)$ with $0\le t\le T$ and $S\subseteq[n]$, determined only by $e_1,\dots,e_T$ and $D$, such that
\[
|\Wit|\le2T\cdot3^\ell,\qquad
1\le|S|\le K\quad\text{for every }(t,S)\in\Wit.
\]
For every trajectory with receiving gap at most $29$ and $M_T\ge D$, some $(t_0,S)\in\Wit$ satisfies
\[
|d_{t_0}(S)|\ge\frac{D^2}{80}.
\]
We call the elements of $\Wit$ witnesses.
\end{lemma}
\begin{proof}
For each $t^*\in\{1,\dots,T\}$ and $i^*\in\{u_{t^*},v_{t^*}\}$, build a rooted tree whose nodes are labeled by triples $(t,S,j)$, with root $(t^*,\{i^*\},0)$. A node is a leaf when any of the conditions
\[
t=0,\qquad S=\emptyset,\qquad |S|=K,\qquad j=\ell
\]
holds. Every other node has the single child $(t-1,S,j)$ if $e_t$ does not cross $S$, and the three children
\[
(t-1,S^0,j+1),\qquad (t-1,S^+,j+1),\qquad (t-1,S^-,j+1)
\]
if $e_t$ crosses $S$, with $S^0,S^+,S^-$ as above.

Time decreases by one along each edge of the tree, so the tree is finite. The set starts with size one, changes in size by at most one at a time, and stops on reaching $K$; thus every node satisfies $|S|\le K$. Along a path from root to leaf, $j$ counts the nodes with three children, so there are at most $\ell$ such nodes. Contracting the edges from nodes with only one child gives a tree of height at most $\ell$ and at most three children per node. It has at most $3^\ell$ leaves. Let $\Wit$ consist of the pairs $(t,S)$ from leaves with $S\ne\emptyset$ in all $2T$ trees. This proves the stated size bounds, and the construction depends only on the edges and $D$.

Now fix a trajectory as in the statement. Let $t^*$ be the first time with $\|d_{t^*}\|_\infty\ge D$. Since $d_0=0$ and only the two endpoints change at each step, each by one, we have $t^*\ge1$ and some endpoint $i^*$ of $e_{t^*}$ satisfies $|d_{t^*}(i^*)|=D$. Replacing all signs by their negatives replaces $d_t,L_t$ by $-d_t,-L_t$ and leaves $2+\varepsilon_tL_t$ unchanged. By~\eqref{eq:lead}, the receiving gap is unchanged, as are $\Wit$ and every $|d_t(S)|$. We may therefore assume that $d_{t^*}(i^*)=D$.

Starting from the root $(t^*,\{i^*\},0)$, we choose a path to a leaf such that every visited node satisfies
\begin{equation}\label{eq:slack}
S\ne\emptyset\qquad\text{and}\qquad\Lambda(t,S)\ge j.
\end{equation}
This holds at the root because $A(1)=D$. At a node with one child, either both endpoints of $e_t$ lie in $S$ or neither does, so $d_{t-1}(S)=d_t(S)$; both the slack and $j$ stay the same. At a node with three children, Lemma~\ref{lem:step} supplies a nonempty child whose slack increases by at least one, while $j$ increases by one. Thus~\eqref{eq:slack} holds along the chosen path.

Let $(t_0,S,j)$ be its leaf. By~\eqref{eq:slack}, $S\ne\emptyset$. If $t_0=0$, then $\Lambda(0,S)=-A(|S|)<0$ by~\eqref{eq:staircase-lower}, contradicting~\eqref{eq:slack}. Hence the leaf satisfies $|S|=K$ or $j=\ell$. In the first case $A(|S|)\ge D^2/80$; in the second case, $j\ge D^2/80$ and $A(|S|)>0$. In either case,
\[
d_{t_0}(S)=A(|S|)+\Lambda(t_0,S)\ge A(|S|)+j\ge\frac{D^2}{80}.
\]
Since this is a leaf with $S\ne\emptyset$, its pair $(t_0,S)$ belongs to $\Wit$.
\end{proof}
Now we are ready to prove the main result.
\begin{proof}[Proof of Theorem~\ref{thm:main}]

Fix the input $e_1,\dots,e_T$ in advance and run the algorithm of Section~\ref{sec:algorithm}. For every fixed $t$, nonempty $S\subseteq[n]$, and $h>0$, Markov's inequality and Proposition~\ref{prop:alg}, with $\theta=\pm\lambda\one_S$, give
\[
\Pr\big(|d_t(S)|\ge h\big)
\le2\inf_{\lambda\ge0}\exp\big(-\lambda h+25\lambda^2|S|\big)
=2\exp\Big(-\frac{h^2}{100|S|}\Big),
\]
where the minimum is attained at $\lambda=h/(50|S|)$.

Let $\Wit$ be the family from Lemma~\ref{lem:witness}. By Proposition~\ref{prop:alg}, every trajectory has receiving gap at most $29$, so
\[
\{M_T\ge D\}\subseteq\bigcup_{(t_0,S)\in\Wit}\Big\{|d_{t_0}(S)|\ge\frac{D^2}{80}\Big\}.
\]
Applying the preceding bound to each witness and using $|S|\le K$, union bound gives
\[
\begin{aligned} 
\Pr(M_T\ge D)
 &\le2|\Wit|\exp\left(-\frac{(D^2/80)^2}{100K}\right)\\
 &\le4T3^\ell\exp\left(-\frac{D^3}{16000}\right)\\
 &\le12T\exp\left(\frac{9D^2}{640}-\frac{D^3}{16000}\right)\\
 &\le12T\exp\left[-\left(\frac D{40}\right)^3\right].
\end{aligned}
\]
Here we used $K\le D/40$, $\ell\le D^2/80+1$, $\log3<9/8$, and
\[
 \frac{9D^2}{640}\le\frac{3D^3}{64000}\qquad(D\ge300).
\] This proves~\eqref{eq:main-tail}.

Set $r=40(c+\log(12T))^{1/3}$ and $D=\max\{300,\lceil r\rceil\}$.
Since $(D/40)^3\ge c+\log(12T)$, the tail bound~\eqref{eq:main-tail} gives
\[
\Pr(M_T\ge D)\le\exp(-c).
\]
Hence, with probability at least $1-\exp(-c)$, $M_T\le D-1 < \max\{300,r\}$, which is~\eqref{eq:main-whp}.

\end{proof}

\section{An inequality in one dimension}\label{sec:onedim}
For an edge $e=\one_u-\one_v$, write a line parallel to $e$ as $x=y+\frac s2e$, where $\langle y,e\rangle=0$. Since $\|e\|^2=2$, the coordinate $s=\langle x,e\rangle=x(u)-x(v)$ changes by $\pm2$ under the moves $x\mapsto x\pm e$. The forcing conditions apply when $|s|>27$, and the coordinate $\langle G,e\rangle$ is centered normal with variance $100$. 

For a finite convex function $g:\R\to[0,\infty)$, define
\begin{equation}\label{eq:Psi}
\begin{aligned}
 F_g(z)&=\min_{|s-z|\le2}g(s),\\
 \kappa_g&=\min_{|\xi|\le27}\max\{g(\xi-2),g(\xi+2)\},\\
 \Psi_g(z)&=\max\{\kappa_g,F_g(z)\}.
\end{aligned}
\end{equation}
Since $F_g\le g$,
\begin{equation}\label{eq:Psi-bounds}
0\le\kappa_g\le\Psi_g\le\max\{\kappa_g,g\}.
\end{equation}We construct a function $\Psi_g$ that bounds the least value attainable by an allowed move and whose Gaussian average is at most that of $g$. Section~\ref{sec:selection} applies this construction to the restrictions $g(s)=f(y+\frac s2e)$ of nonnegative convex test functions.
\begin{figure}[H]
\centering
\includegraphics[width=0.96\textwidth]{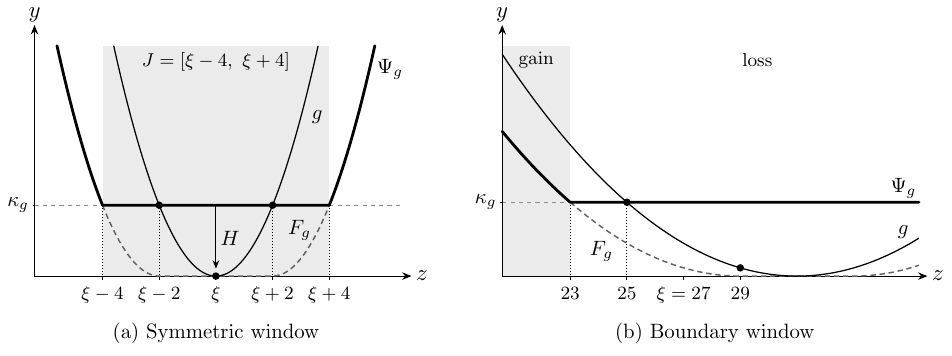}
\caption{The two cases in Lemma~\ref{lem:1d}. The thin, dashed and thick curves represent $g$, $F_g$ and $\Psi_g$, respectively. In (b), the minimizing center is $\xi=27$.}
\label{fig:flattening}
\end{figure}

Continuity of $g$ gives attainment of the minima and continuity of $F_g$ and $\Psi_g$. By convexity, $\kappa_g$ is the smallest level whose sublevel set contains an interval of length $4$ centered in $[-27,27]$.

\begin{lemma}\label{lem:window}
If $\xi$ minimizes the expression defining $\kappa_g$, then
\[
 z\ge\xi\ \Longrightarrow\ g(z-2)\le\Psi_g(z),\qquad
 z\le\xi\ \Longrightarrow\ g(z+2)\le\Psi_g(z).
\]
Thus at least one move has value at most $\Psi_g(z)$, including the forced move when $|z|>27$.
\end{lemma}
\begin{proof}
Choose $s\in[z-2,z+2]$ with $g(s)=F_g(z)$. If $z\ge\xi$, then $\xi-2\le z-2\le s$, so convexity gives
\[
 g(z-2)\le\max\{g(\xi-2),g(s)\}\le\Psi_g(z).
\]
If $z\le\xi$, use $s\le z+2\le\xi+2$ instead. The last assertion follows from $|\xi|\le27$.
\end{proof}

The proof below shows that the loss is smaller than the gain. If the two endpoint values are equal, it is confined to an interval of length $8$, and otherwise it lies in a Gaussian tail.

\begin{lemma}\label{lem:1d}
For $Z\sim N(0,100)$, we have $\E\Psi_g(Z)\le\E g(Z)$.
\end{lemma}
\begin{proof}
Choose a minimizing center $\xi$. Reflection replaces $\xi$ by $-\xi$ and $\Psi_g(z)$ by $\Psi_g(-z)$, so symmetry of $Z$ lets us assume $g(\xi-2)\ge g(\xi+2)$. Write $\slope(a,b)=(g(b)-g(a))/(b-a)$; convexity makes this chord slope nondecreasing in both endpoints.

\emph{Case 1: $g(\xi-2)=g(\xi+2)=\kappa_g$.}
Choose a minimizer $m\in(\xi-2,\xi+2)$ on the window. Such an interior minimizer exists since either the minimum is below the endpoint values or $g$ is constant on the window. Outside the window, chord monotonicity gives $g\ge\kappa_g$, so $m$ is a global minimizer. Put $H=\kappa_g-g(m)\ge0$ and $J=[\xi-4,\xi+4]$.

For $z\in J$, the sliding window meets $[\xi-2,\xi+2]$, so $\Psi_g(z)=\kappa_g$ and $g(z)-\Psi_g(z)\ge-H$. For $z<\xi-4$, the function is nonincreasing on $(-\infty,\xi-2]$, giving $\Psi_g(z)=g(z+2)$, and
\[
 \frac{g(z+2)-g(z)}2
 \le\frac{g(m)-g(\xi-2)}{m-(\xi-2)}\le-\frac H4.
\]
For $z>\xi+4$, the function is nondecreasing on $[\xi+2,\infty)$, so $\Psi_g(z)=g(z-2)$ and
\[
 \frac{g(z)-g(z-2)}2
 \ge\frac{g(\xi+2)-g(m)}{\xi+2-m}\ge\frac H4.
\]
Thus $g(z)-\Psi_g(z)\ge H/2$ on both tails, and
\[
 g(z)-\Psi_g(z)\ge
 \begin{cases}-H,&z\in J,\\H/2,&z\notin J.\end{cases}
\]
Since $J$ has length $8$ and the density of $Z$ is at most $1/(10\sqrt{2\pi})$,
\[
 \Pr(Z\in J)\le\frac8{10\sqrt{2\pi}}<\frac13,\qquad
 \E[g(Z)-\Psi_g(Z)]\ge\frac H2\bigl(1-3\Pr(Z\in J)\bigr)\ge0.
\]

\emph{Case 2: $g(\xi-2)>g(\xi+2)$.}
We prove that $\xi=27$. If not, moving the center to the right strictly lowers the value of the left endpoint by convexity, while continuity keeps the value of the right endpoint below $g(\xi-2)$. Contradiction to minimality. Hence $\kappa_g=g(25)>g(29)$, and $g$ is strictly decreasing on $(-\infty,25]$. Set $\beta=-\slope(23,25)>0$.

For $z<23$, $\Psi_g(z)=g(z+2)$ and chord monotonicity gives $g(z)-\Psi_g(z)\ge2\beta$. For $23\le z\le25$, the window contains $25$, so $\Psi_g(z)=\kappa_g\le g(z)$. For $z>25$, convexity and $
 0\le\Psi_g\le\max\{\kappa_g,g\}$ give
\[
 g(z)-\Psi_g(z)\ge\min\{g(z)-\kappa_g,0\}\ge-\beta(z-25).
\]
Therefore
\[
 \E[g(Z)-\Psi_g(Z)]\ge2\beta\Pr(Z<23)-\beta\E(Z-25)_+\ge0,
\]
because $\Pr(Z<23)\ge1/2$ and $(z-25)_+\le z^2/100$ for every real $z$, whence $\E(Z-25)_+\le1$.
\end{proof}
\section{Proof of the selection theorem}\label{sec:selection}
Fix $e=\one_u-\one_v$ and write $x^\perp=x-\frac12\langle x,e\rangle e$ for the projection onto $e^\perp$. Now we prove the high dimensional version of Lemma~\ref{lem:1d}.

\begin{lemma}\label{lem:majorant}
For $f\in\mathcal C_+$, let $g_y(s)=f(y+\frac s2e)$ for $y\in e^\perp$ and put
$\Psi_f(x)=\Psi_{g_{x^\perp}}(\langle x,e\rangle)$. Then $\Psi_f\in\mathcal C_+$,
\begin{equation}\label{eq:majorant-integral}
 \int\Psi_f\,d\gamma\le\int f\,d\gamma,
\end{equation}
and for every $x$ there is $q\in B_x$ such that
\begin{equation}\label{eq:majorant-move}
 qf(x+e)+(1-q)f(x-e)\le\Psi_f(x).
\end{equation}
\end{lemma}
\begin{proof}
Define
\[
\begin{aligned}
 F_f(x)&=\min_{|r|\le1}f(x+re),\\
 K_f(y)&=\min_{|\xi|\le27}\max_{\varepsilon=\pm1}
 f\bigl(y+(\xi/2+\varepsilon)e\bigr),\qquad y\in e^\perp.
\end{aligned}
\]
Then $F_f(x)=F_{g_{x^\perp}}(\langle x,e\rangle)$, $K_f(y)=\kappa_{g_y}$, and
$\Psi_f(x)=\max\{F_f(x),K_f(x^\perp)\}$.
Both objectives are jointly convex in the point and the scalar variable, and the minima are attained over fixed compact intervals. Taking convex combinations of minimizers proves convexity: for minimizers $r_0,r_1$ at $x_0,x_1$ and $\rho\in[0,1]$,
\begin{align*}
 F_f((1-\rho)x_0+\rho x_1)
 &\le f\bigl((1-\rho)(x_0+r_0e)+\rho(x_1+r_1e)\bigr)\\
 &\le(1-\rho)F_f(x_0)+\rho F_f(x_1).
\end{align*}
The same argument applies to $K_f$. Thus $\Psi_f$ is finite, nonnegative and convex.

Since $f(x\pm e)=g_{x^\perp}(\langle x,e\rangle\pm2)$, Lemma~\ref{lem:window} gives~\eqref{eq:majorant-move}, with $q=0$ or $1$ respecting the forcing conditions.

Finally, let $Y=G^\perp$ and $Z=\langle G,e\rangle$. Orthogonal Gaussian components are independent, $Z\sim N(0,100)$, and $G=Y+\frac12Ze$. Nonnegativity, Tonelli's theorem and Lemma~\ref{lem:1d} give
\[
 \int\Psi_f\,d\gamma
 =\E_Y\E_Z\Psi_{g_Y}(Z)
 \le\E_Y\E_Zg_Y(Z)
 =\int f\,d\gamma<\infty.
\]
This proves~\eqref{eq:majorant-integral} and $\Psi_f\in\mathcal C_+$.
\end{proof}

Lemma~\ref{lem:majorant} handles one test function at a time. Next we prove a finite affine separation lemma which combines finitely many into one function.
\begin{lemma}\label{lem:affine-separation}
Let $B\subset\R^N$ be a nonempty compact convex set, and let
$\varphi_1,\ldots,\varphi_m:\R^N\to\R$ be affine functions. Suppose that
\[
B=\bigcup_{i=1}^m\{p\in B:\varphi_i(p)>0\}.
\]
Then there exist $\lambda_1,\ldots,\lambda_m\ge0$ such that
\[
\sum_{i=1}^m\lambda_i\varphi_i(p)>0
\qquad\text{for every }p\in B.
\]
\end{lemma}
\begin{proof}
The set $K=\varphi(B)$, where $\varphi=(\varphi_1,\dots,\varphi_m)$, is compact and
convex, and by assumption it is disjoint from the closed convex set $Q=(-\infty,0]^m$.
By the Hahn-Banach separation theorem there are $\lambda\in\mathbb R^m$ and $a<b$ with
$\langle\lambda,y\rangle\le a$ for $y\in Q$ and $\langle\lambda,z\rangle\ge b$ for $z\in K$.
Taking $y=-s\mathbf 1_i$ and letting $s\to\infty$ gives $\lambda_i\ge0$; taking $y=0$
gives $a\ge0$. Hence $\sum_i\lambda_i\varphi_i(p)=\langle\lambda,\varphi(p)\rangle\ge b>0$
for every $p\in B$.
\end{proof}

Now we complete the argument by reducing to a single test function.

\begin{proof}[Proof of Theorem~\ref{thm:selection}]
Recall from Section~\ref{sec:admissible} that $\mathcal B$ is nonempty, compact and convex. For $f\in\mathcal C_+$ and $p\in\mathcal B$, let
\[
\begin{aligned}
\Phi_f(p)&=\int f\,d\mu^{p,e}-\int f\,d\gamma\\
&=\sum_{x\in\supp\mu}\mu(\{x\})\big(p(x)f(x+e)+(1-p(x))f(x-e)\big)-\int f\,d\gamma.
\end{aligned}
\]
The formula extends $\Phi_f$ to an affine function on $\R^{\supp\mu}$, so $\mathcal B_f=\{p\in\mathcal B:\Phi_f(p)\le0\}$ is closed and convex. By the criterion in Section~\ref{sec:admissible}, the set of admissible selections is $\bigcap_{f\in\mathcal C_+}\mathcal B_f$. It is therefore compact and convex. To prove nonemptiness, it suffices to establish the finite intersection property.

First fix a single $f\in\mathcal C_+$.~\eqref{eq:majorant-move} gives, for each $x\in\supp\mu$, a choice $p(x)\in B_x$ with $p(x)f(x+e)+(1-p(x))f(x-e)\le\Psi_f(x)$. These choices form a selection $p\in\mathcal B$, and
\[
\Phi_f(p)\le\int\Psi_f\,d\mu-\int f\,d\gamma
\le\int\Psi_f\,d\gamma-\int f\,d\gamma\le0.
\]
Here the second inequality uses $\Psi_f\in\mathcal C_+$ and $\mu\cx\gamma$, and the last is~\eqref{eq:majorant-integral}. Thus every single test has a selection satisfying its inequality.

Now let $f_1,\ldots,f_m\in\mathcal C_+$ and suppose that
$\bigcap_{i=1}^m\mathcal B_{f_i}=\varnothing$. Then
\[
\mathcal B=\bigcup_{i=1}^m\{p\in\mathcal B:\Phi_{f_i}(p)>0\}.
\]
Applying Lemma~\ref{lem:affine-separation} with $B=\mathcal B$ and
$\varphi_i=\Phi_{f_i}$ gives $\lambda_1,\ldots,\lambda_m\ge0$ such that
$\sum_{i=1}^m\lambda_i\Phi_{f_i}$ is strictly positive on $\mathcal B$.
Consequently,
\[
f=\sum_{i=1}^m\lambda_i f_i\in\mathcal C_+,
\qquad
\Phi_f(p)=\sum_{i=1}^m\lambda_i\Phi_{f_i}(p)>0
\quad(p\in\mathcal B),
\]
contradicting the preceding conclusion for this $f$. Thus the closed sets
$\mathcal B_f$ have the finite intersection property. Therefore, the set of admissible selections is nonempty.
\end{proof}

\begin{remark}\label{rem:banaszczyk}
The function $\Psi_f$ is a convex majorant of the minimum over allowed one-step moves:
\[
 \min_{q\in B_x}\bigl\{qf(x+e)+(1-q)f(x-e)\bigr\}\le\Psi_f(x).
\]
Its relation to Banaszczyk's transform~\cite{Ban98} can be seen on sublevel sets. For $y\in e^\perp$ and $a\ge0$, let $I_{y,a}=\{s:g_y(s)\le a\}$, an interval that may be empty or unbounded. The definitions give
\[
 \{z:\Psi_f(y+\tfrac z2e)\le a\}
 =\begin{cases}
 I_{y,a}+[-2,2],&\text{if }[\xi-2,\xi+2]\subseteq I_{y,a}
                  \text{ for some }|\xi|\le27,\\
 \varnothing,&\text{otherwise}.
 \end{cases}
\]
Thus each sublevel interval is expanded by $2$ at both ends if it contains an allowed segment, and is discarded otherwise. Banaszczyk's transform similarly expands the sufficiently long sections of a convex set and discards the shorter ones; here the segment centers are additionally restricted to the slab $\{x:|\langle x,e\rangle|\le27\}$ to enforce the prescribed moves. Equation~\eqref{eq:majorant-integral} gives the corresponding Gaussian integral inequality.
\end{remark}

\section{AI disclosure}
GPT 6 Astra first proposed the approach via properties (P1) and (P2) and the functional analogue of Banaszczyk’s transform. 
Claude Opus 5.5 was used to review drafts, suggest changes, and to assist with preparing the illustrations for the manuscript.
The author determined the explicit constants and notations used in the theorem; identified and corrected errors; verified every proof; simplified the exposition and wrote the paper.
The author takes full responsibility for the contents of this paper.

\section{Acknowledgment}
The author is deeply grateful to Professor Konstantin Tikhomirov, who introduced him to this problem and has followed the work closely from the first rough idea to the final draft. He asked questions exposing a gap in what the author thought was complete and pointing to a simpler route. In particular, the simplification in proving Theorem 3.3 is due to him. 

\providecommand{\etalchar}[1]{$^{#1}$}

\end{document}